\documentstyle [aps,prl,epsf,twocolumn] {revtex}

\begin{document}

\renewcommand{\[}{\begin{equation}}
\renewcommand{\]}{\end{equation}}


\title{Glass Model, Hubbard Model and High-Temperature Superconductivity}
\author{Ingo Morgenstern, Werner Fettes, Thomas Husslein,\\
	  {\it	 Fakult\"at f\"ur Physik, Universt\"at Regensburg, 93053 Regensburg, Germany}\\
 Dennis M. Newns, and Pratap C. Pattnaik\\
{\it IBM T.J. Watson Research Center, P.O. Box 218, Yorktown  Heights, NY 10598, USA} }
\date{\today}

\maketitle

\begin{abstract}
In this paper we
revisit the glass model describing the
macroscopic
behavior of the High-Temperature superconductors. We link the glass model
at the microscopic level
to the striped phase phenomenon, recently discussed widely.
The size of the striped phase domains is consistent with
earlier predictions
of the glass model when it was introduced for High-Temperature
Superconductivity in 1987.
In an additional step we use the Hubbard model to describe the microscopic
mechanism for d-wave pairing within these finite size stripes.

We discuss the implications for superconducting correlations of 
Hubbard model, which are
much higher for stripes than for squares, for finite size scaling, 
and for the new view of the glass model picture.
\end{abstract}

\bigskip


\section{Introduction and Glass Model}

More then 10 years have passed since the introduction of the glass model in
High-Temperature Superconductivity 
\cite{MOR87,SCH87/2,MOR88/4,PUE88,MOR88/5,KEL88,MOR89/2,ROS89}.
The model was originally
used for the description of granular superconductors \cite{EBN85}.
But it became immediately clear that the glass behavior of the then new
High-$T_c$ materials was not caused by the granular structure of the
ceramic compounds; rather the magnetic fields
used in the experiments suggested that the microscopic
origin of glassy behavior occurred at a length scale smaller than that of
the grains \cite{MOR87}.

In this paper we revisit the glass model. Inspired
by the recent study by Tsuei and Doderer \cite{TSU98} of the charge confinement effect
in cuprate superconductors,
we link the intragrain inhomogeneities demanded in the glass
theory to the microscopic
concept of striped phase \cite{Tranq,Bianc}, and in particular to that of
stripe {\it domains}
\cite{CHE89,TSU98} which have
recently become of wider interest to the High-$T_c$ community.
The stripe domains are identified with the 
areas of constant phase in the glass theory.
Microscopically, Tsuei and Doderer used their charge confinement
model \cite{TSU98}
to explain various generic features of the normal-state  pseodogap,
including the magnitude of the gap.
Such stripe domains may be described from first principles by a model such
as the Hubbard Model, enabling an ab initio numerical calculation of
intra-domain pairing properties such as the effective pairing temperature
and pseudogap.

In \cite{MOR87} a behavior reminiscent of spin glasses was detected,
the quasi-de-Almeida-Thouless-line (quasi AT-line):
\begin{equation}
\label{eqatline}
H^{2/3} \approx T_c(H) - T_c(0)
\quad ,
\end{equation}
where $H$ is the magnetic field and $T_c(H)$ the corresponding critical temperature
or to be more precise the temperature below which metastable behavior
occurs as seen in \cite{MOR89/2}. The existence of the AT-line
was first seen as an evidence for glass behavior, but it turned out, 
that a totally different physical mechanism was the cause for the 
$\rm H^{2/3}$ behavior \cite{MOR87}. Numerical simulations of the
High-$T_c$ glass model \cite{MOR87} lead to this conclusion.

The glass model can be described by the Hamiltonian \cite{MOR87}
\begin{eqnarray}
\label{eqhglass}
{\cal H} = - J \sum_{<i,j>} \cos\left(\phi_i - \phi_j -
A_{ij}\right)
\quad ,
\end{eqnarray}
where
\begin{eqnarray}
\label{eq3}
A_{ij} = \int_i^j {\vec A} dl
\quad .
\end{eqnarray}
$\vec A$ is the vector potential leading to \cite{MOR87}
\begin{eqnarray}
\label{eqphases_2}
A_{ij}=\frac{2\pi}{\phi_0} H \frac{x_i+x_j}{2}(y_j-y_i)
\quad .
\end{eqnarray}
In equation (\ref{eqhglass}) to (\ref{eqphases_2})
$\phi_0$ is the elementary flux quantum, $H$ the magnetic field.
$\phi_i$ is the phase of the
superconducting wave function in an area or domain denoted by i; $x_i$ and $y_i$ are the
cartesian coordinates. Furthermore $J$ is the coupling energy
between two clusters according to \cite{EBN85} taken identical
in the simulations in \cite{MOR87}, $\int_i^j$ the line integral from
cluster $i$ to $j$, $<i,j>$ the sum over the nearest neighbors,
and finally $A_{ij}$ are the phase factors between cluster $i$ and $j$.

\begin{figure}[hbtp]
\begin{center}
\begin{minipage}{6.0cm}
\epsfxsize 6.0cm \epsfbox{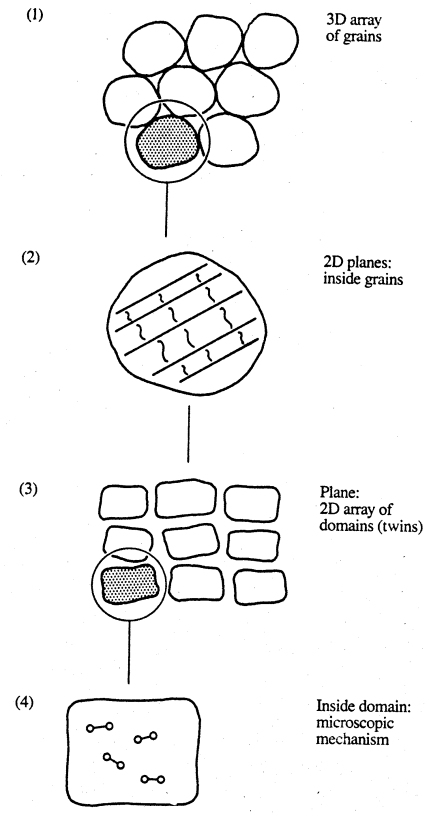}
\end{minipage}
\caption{\label{fig1}Schematic illustration of the physical situation
in a single crystal of a High-$T_c$ ceramic.
(analogous to figure 1 in $[$7$]$)
}
\end{center}
\end{figure}

As a result of the simulations in \cite{MOR87} we consider 
figure~\ref{fig1} which is reproduced
from \cite{MOR87} and follow-up papers. Because of the results of the 
simulations we had to conclude, that the origin of the glass behavior 
lies inside the grains, therefore also inside single crystals.

The $H^{2/3}$ behavior was only reproduced in two dimensions,
therefore the domains or areas of constant phase $\phi_i$ had to
be located inside the CuO-planes, connected by weak links $J$
(figure~\ref{fig1}).  The size of the domains was also known.
Here we present a typical calculation of estimating 
the domain size with data
available at the time of the publication of \cite{MOR87,MUE87}.

\begin{figure}[hbtp]
\begin{center}
\begin{minipage}{7.0cm}
\epsfxsize 7.0cm \epsfbox{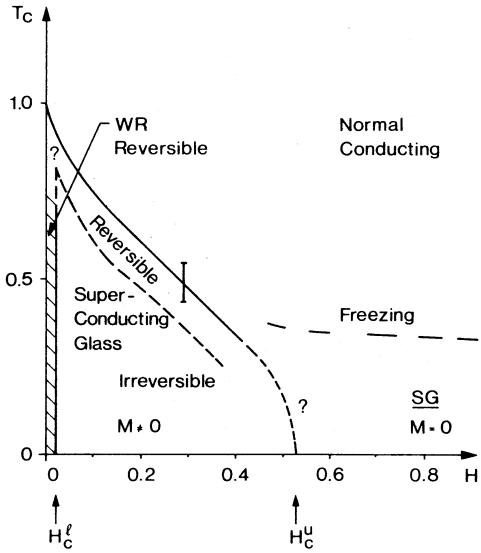}
\end{minipage}
\caption{\label{fig2} Phase diagram in $H-T$ plane. Reversible
weak-random (WR) phase between $H\approx 0.0$ and $H^l_c \approx
0.01$ - $0.03$. Between $H^l_c$ and $H^u_c \approx 0.5$,
superconducting glass phase. Irreversible effects are separated
from reversible effects by dashed line. Above $H^u_c$, usual
XY--spin glass phase (SG). Dashed line indicates freezing
transition.
(analogous to figure 10 in $[$1$]$)
}
\end{center}
\end{figure}

As the magnetic field of reference we consider the upper critical
field $H_c^u$. In figure~\ref{fig2} we see $H_c^u$
in the phase diagram of equation (\ref{eqatline}), that $H_c^u\approx 0.5$.
We should denote the ${2\pi}/{\phi_0}$ was taken as unity as
is normal in numerical simulations.

\begin{figure}[hbtp]
\begin{center}
\begin{minipage}{8.0cm}
\epsfxsize 8.0cm \epsfbox{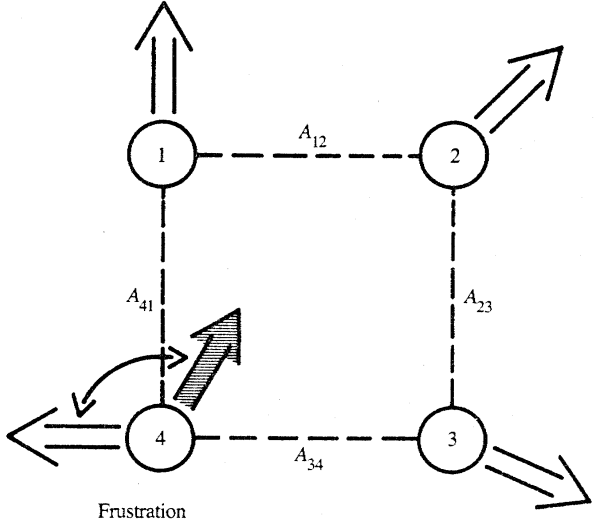}
\end{minipage}
\caption{\label{fig3}
Plaquette with domains of constant phase $\phi_i$ and phase differences
$A_{ij}$ (see text)
}
\end{center}
\end{figure}

For an estimation of the typical distance $a$ between two clusters, we
consider a plaquette in figure~\ref{fig3}.

In the simulation $a = 1$.
The idea is that the phase factors are equal, $A_{ij, sim} = A_{ij, exp}$, for
corresponding fields $H$. ($A_{ij, sim} $ denoting $A_{ij}$ used in
simulation and $A_{ij, exp}$ correspondingly in experiment).
Therefore for $H^u_c$ we have
$A_{ij, sim} = 0.5$.
Considering the experimental situation we refer to figure~1
in \cite{MUE87}. There the magnetization or magnetic moment
$\mu(H)$ is plotted
versus $H$ for various temperatures. Considering the lowest
temperature $\mu(H)$ tends to zero at $ 1.5 T$.
Therefore $H_c^u \approx 1.5 T $ (Tesla) in experiment with
$\phi_0 = 2.06 \cdot 10^{-15} m^2 T$
we find that $a^2 \approx 1 \cdot 10^4 {\rm \AA}^2$ and
therefore $a \approx 100 {\rm \AA}$.

We wish to emphasize here, that the length scale of
$100 {\rm \AA}$ deduced from the glass picture is consistent with the
length of the stripe domains seen in STM data \cite{CHE89},
and roughly with that of the stripe width determined from
neutron\cite{Tranq} and ultraviolet
photoemission\cite{Bianc} data.  We propose revision of the concept of
clean, single crystal high temperature superconductor material
to include the concept of striped phase--type inhomogeneity with
associated `glassy' behavior.
It should be noted, that the aging or memory effect in spin glasses
was also detected \cite{ROS89}.

At the end of this section we would like to counter a certain source of
criticism concerning the glass model. Ten years ago the community only
considered the case of s-wave superconductivity. In the model $J=1$ was
chosen. The glass behavior as described in \cite{MOR87} is totally
dominated by frustration, which is caused by the $A_{ij}$'s in equation
(\ref{eqhglass}). In
s--wave superconductivity $J$ would not change sign and therefore
not lead to additional frustration. This could be different in d--wave
as pointed out by Sigrist and Rice \cite{SIG92}.

But the situation in
the simulation of \cite{MOR87} was, that only weak ''correlated'' disorder was
chosen in the framework
of the square lattice. This means the coordinates
$x_i$ and $y_i$ in equation (\ref{eqphases_2}) were obtained by allowing
a random displacement of the sites within a circle with radius $r\approx 0.4$
of the lattice constant around the original site of the square lattice.
Therefore $J$ does not change sign even in the case of d-wave
superconductivity. It should be pointed out that only this type of
weak ''correlated'' disorder \cite{MOR87} in two dimensions reproduced
the $H^{2/3}$ behavior of the quasi AT-line \cite{MUE87}. At that time this
result also lead to the prediction of a roughly two-dimensional
plane-like anisotropy in the new superconductors \cite{MOR87}.

Summarizing the glass model picture of \cite{MOR87} is still
applicable to the correlated striped domains in the High-$T_c$ 
materials governing the d-wave symmetry and predicting  
the domain sizes correctly.

\section{Microscopic Mechanism and Hubbard Model}

As pointed out the microscopic mechanism has to be considered inside the
domains. While glass papers did not deal with the mechanism, it was clear
at that time, that, if the glass picture applies, the observed $T_c^{exp}$
is $T_c^{glass}$ of the glass model. The microscopic mechanism could easily have a
much higher critical temperature $T_c^{m}$.

Inside the domains we propose an electronic mechanism as suggested by
the Hubbard model (HM) or to be more precise the tt'--HM 
\cite{MOR94,HUS94}
\[
{\cal H} =
{\cal H}_{kin} + {\cal H}_{pot}
\]
with the kinetic
\[
{\cal H}_{kin} =
- \sum_{i,j, \sigma} t^{}_{i,j} (c^{\dagger}_{i,\sigma} c^{}_{j,\sigma}
 +c^{\dagger}_{j\sigma} c^{}_{i\sigma})
\]
and the potential part
\[
{\cal H}_{pot} =
U \sum_{i} \left(n_{i, \uparrow}-\frac{1}{2}\right)
\left( n_{i, \downarrow} - \frac{1}{2} \right)
\label{glei1}
\]
of the Hamiltonian.  We denote the creation operator for an electron with
spin $\sigma$ at site  $i$ with $c^\dagger_{i,\sigma}$, the corresponding
annihilation operator with $c^{}_{i,\sigma}$, and the number operator at site
$i$ with $n_{i,\sigma} \equiv c^\dagger_{i,\sigma} c^{}_{i,\sigma}$. $t_{i,j}$
is the hopping parameter only nonzero between nearest neighbors $i,j$
($t_{i,j}=t$) and next nearest sites $i,j$ ($t_{i,j}=t'$),
and finally $U$ is the interaction being repulsive in the
High-$T_c$ case.

The sites of the HM can be associated with the copper in the
CuO-layers of the High-$T_c$ crystals \cite{AND87}.

We published evidence for superconducting behavior with d-wave symmetry
of the repulsive HM as early as 1994 \cite{MOR94}.
The d-wave symmetry in the superconducting state
was subsequently established
for the High-$T_c$ materials \cite{TSU94}.

According to \cite{SCA89,FRI91}  we use the (vertex)
correlation function (instead of the largest eigenvalue)
to provide evidence for superconductivity following the
standard concept of off diagonal long range order \cite{YAN62}.
The vertex correlation function (vertex CF) is defined for
d$_{x^2-y^2}$-waves as
\[
\label{EqVertexdxmy}
C^V_{d}(r) = C_d(r) -
\frac{1}{L} \sum\limits_{i,\delta,\delta'} g_\delta
g_{\delta'}
C^{}_\uparrow (i,i+r) C^{}_\downarrow (i+\delta,i+r+\delta')
\]
with the single particle correlation function
$C^{}_\sigma\equiv \langle c^\dagger_{i,\sigma} c^{}_{i+r,\sigma}\rangle$
for spin $\sigma$
to extract the pairing effects in the two particle CF
\[
\label{Eqfulldxmy}
C_{d}(r) =
\frac{1}{L}
\sum\limits_{i,\delta,\delta'} g_\delta g_{\delta'}
\langle c_{i\uparrow}^\dagger c_{i + \delta \downarrow}^\dagger
c_{i+r + \delta' \downarrow}^{} c_{i+r\uparrow}^{}
\rangle
\]
with the phase factors $g^{}_\delta$, $g'_\delta= \pm 1$ to model the
d$_{x^2-y^2}$--wave symmetry, the number of lattice points $L$ and
the sum $\delta$ ($\delta'$) over all nearest neighbors.

\begin{figure}[hbtp]
\begin{center}
\begin{minipage}{7.0cm}
\epsfxsize 7.0cm \epsfbox{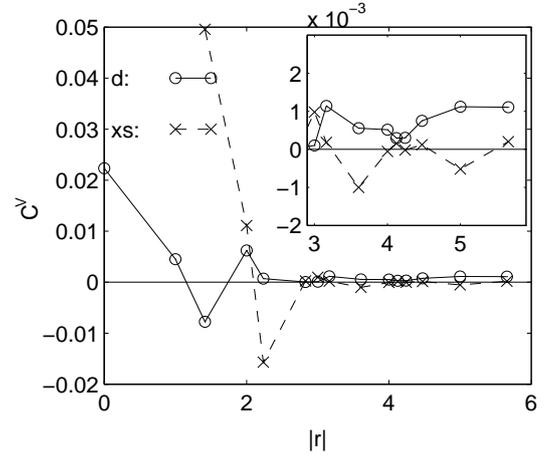}
\end{minipage}
\caption{\label{fig4}
Vertex-CF versus Cooper pair distance $|r|$ for extended s-wave ($xs$)
and d$_{x^2-y^2}$-wave ($d$) for a $8\times 8\times 1$ system with
$U=2$, $t'=-0.22$ and $\langle n \rangle = 0.78$.
(Inset shows the long range regime enlarged.)
}
\end{center}
\end{figure}

In figure~\ref{fig4} we reproduce the evidence for d-wave pairing
in the repulsive HM using the projector quantum Monte Carlo (PQMC) 
technique.  These simulations
\cite{SUG86,SOR89} performed for the relatively
weak--coupling value of $U=2$, were found, with adequate numerical
precautions, to be quite stable \cite{HUS97,FET98/3,FET97/3}. 
For higher $U$--values, e.g.
$U=4$, PQMC is found to encounter difficulties for larger than
$8\times 8$ cluster sizes, and in fact evidence for finite plateaus was
not reproduced, e.\,g.\ \cite{HIR88,WHI89,IMA91}, 
for large $U$. However in the simulation technique of
Zhang, Carlson and Gubernatis
\cite{ZHA97} the existence of the plateau phenomenon in the
vertex CF (figure~\ref{fig4}) was confirmed.  Recent Variational Monte Carlo
results \cite{YAM98} also confirm the existence of d-wave pairing in Hubbard clusters.

Given the observation of plateaus supporting the existence of d--wave
pairing in finite clusters, what are the implications for bulk
superconductivity? Up to now it was thought necessary to scale the
finite cluster calculations to infinite size in order to answer this
question, which unfortunately presents difficulties as
corrections to scaling are tremendously large compared to
classical systems \cite{HUS96,FET98/3,FET98}.
However in view of the new stripe phase picture of the crystalline
High-$T_c$ material, the simulated cluster can be reinterpreted as a single
finite length stripe as seen in STM data \cite{CHE89}, with typical
dimensions of roughly $12 \times 4$
unit cells, i.e. even smaller
than our largest system sizes ($16\times 16$ in \cite{FET98}).
Scaling to an infinite system is inappropriate as the PQMC cluster
calculation is now descriptive of the actual quantum systems
present in the
crystal (although they are treated as totally isolated).

\begin{figure}[hbtp]
\begin{center}
\begin{minipage}{7.0cm}
\epsfxsize 7.0cm \epsfbox{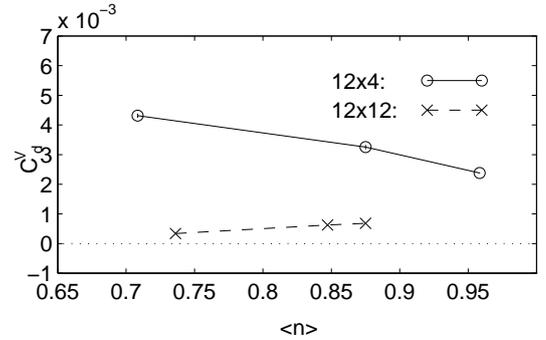}
\end{minipage}
\caption{\label{fig1204}
Averaged ($|r_c|=1.9$) vertex CF versus filling for the tt'--HM
with $U=2$, $t'=-0.22$ and the system sizes $12\times 4$, and
$12\times 12$. The parameter of the PQMC were $\theta =8$ and $\tau=0.125$.
Lines are to guide the eye.
}
\end{center}
\end{figure}

We consequently performed additional simulations for $12\times 4$
systems and now report the remarkable result that the plateau
($4\times 12$) is about a factor of five higher than the plateau
($12\times 12$), figure~\ref{fig1204}. We averaged only the vertex
CF $C^V_d(r)$ in the large range regime of $r$, i.\,e.\ for the
distances $|r|>|r_c|$:
\[
\bar{C}^{V,P}_d \equiv	\frac{1}{L_c} \sum_{r,|r|>|r_c|} C^v_d(r)
\]
with the number $L_c$ of points with $|r| > |r_c|$.
The qualitative behavior of our results is not influenced by the
choice of $r_c$ as long as we suppress the short range correlations
(i.\,e.\ $r_c\ge 1.9$).

Due to finite size effects (see \cite{FET97/2,FET98} or in more
detail \cite{FET98/3}) the shape of the curves in respect to the
filling in figure~\ref{fig1204} is influenced by $\langle n\rangle$ and
$L$.

\begin{figure}[hbtp]
\begin{center}
\begin{minipage}{6.5cm}
\epsfxsize 6.5cm \epsfbox{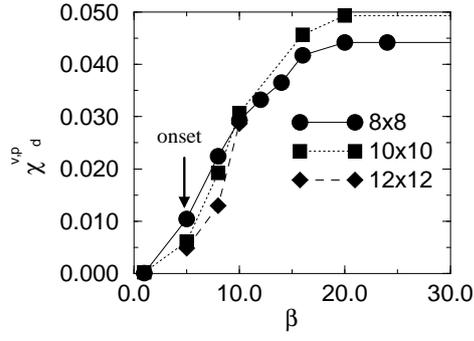}
\end{minipage}
\caption{\label{figgqmc}
Cumulated d-wave Vertex-CF $\chi^{V,P}_d$ versus inverse
temperature $\beta$ for the
two-dimensional Hubbard model with $U=2$, $t'=-0.22$ and
$\langle n\rangle \approx 0.8$.
(data from figure 5.10 in $[$33$]$.)
}
\end{center}
\end{figure}

Considering the estimates of the microscopic critical temperature
$T_c^{m}$ from grand canonical
Monte Carlo (GQMC) \cite{BLA81} simulations for the squared
finite Hubbard model in \cite{HUS96}
we can calculate $T_c^{m}$ from the
$\chi^{V,P} \equiv L \bar{C}^{V,P}_d$ versus $\beta$ plot. ($\beta$ is
the inverse temperature.)
In figure~\ref{figgqmc} we reproduce the results of the  GQMC
runs in \cite{HUS96}. We take as estimate of $T_c^m$ the inflection point
of the $\chi^{V,P}_d$ versus $\beta$ curve. From figure~\ref{figgqmc}
we deduce $\beta$ to be roughly $\beta \approx 10$.
Taking the hopping parameter $t=0.1$\,eV this corresponds to the
temperature $T_c \approx 100K$.

In figure~\ref{figgqmc} all curves
for different system sizes intersect at one temperature. Even so
the magnitude of the superconducting signal at low temperatures
is increased for striped geometries, we expect from figure~\ref{figgqmc},
that the value of $T_c^m$ should be approximately the same. 
Here "$T_c$" denotes the temperature where the
correlation length equals the size of the systems.

The first occurrence of preformed pairs in the finite Hubbard clusters
is at an even higher temperature $T^*$.
$T^*$ is at the onset of the susceptibility, i.e.\ the temperature
with the first occurrence of a nonzero susceptibility, in
figure~\ref{figgqmc}. A rough estimate is $\beta \approx 5$.
With $t = 0.1$\,eV this corresponds to $T^*\approx 200$\,K, which is in
agreement with experimental results 
\cite{DIN96,LOE96,LOE96/2,HAR96,KELpriv97}.
It should be noted, that the range for $t$ is between $0.1$\,eV to 1\,eV
as deduced from the experiments \cite{HYB89,BAC91,SON95}, thus this onset
could be even at higher temperatures \cite{DENpriv}.

\begin{figure}[hbtp]
\begin{center}
\begin{minipage}{6.5cm}
\epsfxsize 6.5cm \epsfbox{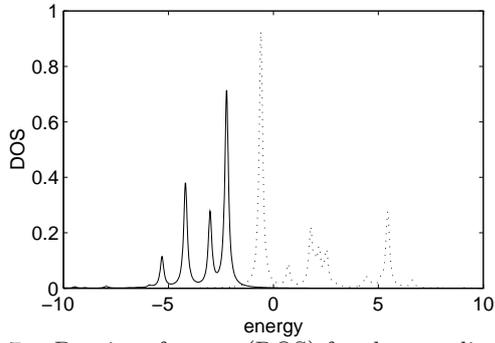}
\end{minipage}
\caption{\label{figspectr}
Density of states (DOS) for the
two-dimensional Hubbard model with $U=2$, $t'=-0.22$ and
$\langle n\rangle \approx 0.75$  and $L=8$.  Energy axis in units of $t$.
(data from $[$46$]$)
}
\end{center}
\end{figure}

Recently Tsuei and Doderer \cite{TSU98}, have pointed out a possible
interpretation of the pseudogap as a function of doping in
cuprates as the finite size
gap in the quasiparticle spectrum of the stripe phase domains.
Consistent with these ideas, we also
identify the onset temperature $T^*$ with the pseudogap energy in
the cuprate material.
Numerical values for the energy gap
between the ground state and the first excited state in finite
size cluster systems were obtained already in
1995 \cite{FET95}.  Figure~\ref{figspectr} reproduces a typical case
for an $\sqrt{8} \times \sqrt{8}$ system as seen in Lanczos
diagonalization (the finite size gap is large for the relatively small clusters
for which Lanczos diagonalization is tractable).
The experimentally seen superconducting gap
is different from the pseudogap (finite size gap).
It is the gap seen in the macroscopic glass model
when the system as a whole becomes superconducting and the finite gaps
are correlated. Therefore it is clear that for the superconducting
(glass model) gap a clear d-wave symmetry is seen \cite{TSU94}
while in the case of the pseudogap the effect is smeared out
\cite{LOE96,HAR96} as the experiment
''averages'' over different (finite size) clusters and slight
correlated disorder.

\section{Summary and Conclusions}

The new view of High-$T_c$
superconductivity emerging from Ref.\ \cite{TSU98} and the present paper
is based macroscopically on the
interaction of nanoscopic, internally-paired, domains (glass model)
and microscopically on the enhancement of pairing found to occur 
within metallic domains constituting the
striped phase.  The glass model for example enables us to identify
the superconducting $T_c$ with the $T_c^{glass}$ of the glass model.

The microscopic model of a domain can be exploited, based on a Hamiltonian
such as the Hubbard model, to estimate the relatively high
temperature $T^{*}$, the effective d-wave
intra--domain pairing temperature, and to determine the pseudogap which
is
identified \cite{TSU98} with the finite size quasiparticle gap within the domain.
In  earlier simulations we already found the
first signs of pairing in the Hubbard model at remarkably 
high temperatures \cite{HUS96}. We find in the present work
a very large enhancement of the intra--domain pairing when a
non-square cluster morphology,
$12\times 4$, appropriate for the striped phase,
is considered.

A further consequence of the glass model is that all
changes in $T_c^{exp}$ are related to the interdomain
coupling energy $J$ of equation~(\ref{eqhglass})
and not to the microscopic mechanism described by the Hubbard model.
This means, e.g.\ for the isotope effect, that when masses are
altered by changing $\rm O^{16}$ to $\rm O^{18}$, leading to
the isotope exponent $\alpha = {\Delta T_c}/{\Delta M}$, the mass
effect is entering through an isotope shift
in $J$ or other glass model parameters. $T^*$ is probably not
affected by the isotope effect \cite{KELpriv}. This would be another
evidence for the combined glass model -- Hubbard model picture. 

In general we note that all experiments aiming at changing $T_c^{m}$ and
therefore the microscopic mechanism have to be reinterpreted or
redone in terms of $J$ of the glass model.
Future research
could usefully concentrate on these weak links as seems to be
occurring \cite{KIV98}. Only the glass model $J$ governs
the $T_c^{exp}$ (in fact $T_c \approx J$), and additionally 
the critical currents of the materials.
Progress in increasing $T_c$ or the critical current so far has been
only accidental as the underlying glass behavior was not recognized.

In the scientific discussion about ten years ago it was concluded 
\cite{MOR89/2} that the glass model would be only applicable to so 
called "bad" or "glassy" samples, i.e.\ samples with weak links 
described by Hamiltonian of equation (\ref{eqhglass}) \cite{MOR89/2}. 
At this time future research had to concentrate on "good" samples 
without weak links. Here we argue again, that there are no "good"
samples, as the $T_c^{exp}$'s for all samples, single crystals, thin
films, and ceramics, are for a definite composition about the same. 
It follows consequently that the $T_c^{exp}$ is the $T_c^{glass}$ 
of the glass model.

Furthermore it is clear as we deal with finite size domains (or
stripes) that mean field theory and therefore BCS does not apply.

Future research could usefully emphasize
the origin and manipulation of the weak links (i.\,e.\ the
$J$ in the glass model) between the domains (or stripes).

\section{Acknowledgment}

The authors would like to express their thanks to C.C. Tsuei and T. Doderer
for making available a preprint of their work \cite{TSU98}.
We would like to acknowledge very helpful discussions with 
T. Schneider, H. Keller, D. Brinkman, Ch.\ Rossel, J.G. Bednorz, 
J.P. Loquet, H. De Raedt and U. Krey. Especially we would like to thank
K.A. M{\"u}ller for inspiring discussions and ideas.
In addition, W. F. is grateful for the financial support of the DFG
(Deutsche Forschungsgemeinschaft).
A part of the numerical simulations were performed on the SP2
parallel computer of the Leibnitz Rechenzentrum Munich, which
grants us a generous amount of CPU time.
Finally we acknowledge the financial support of the UniOpt GmbH,
Regensburg.


\bibliographystyle{unsrt}

\begin{thebibliography}{10}

\bibitem{MOR87}
I.~Morgenstern, K.A. M{\"u}ller, and J.G. Bednorz.
\newblock {\em Z. Phys.}, {\bf B69}, 33 (1987).

\bibitem{SCH87/2}
J.W. Schneider, Hp. Baumeler, H.~Keller, W.~Obermatt, B.D. Patterson, K.A.
  M{\"u}ller, J.G. Bednorz, K.W. Blazey, I.~Morgenstern, and I.M. Savic.
\newblock {\em Phys.\ Letters}, {\bf A124}, 107 (1987).

\bibitem{MOR88/4}
I.~Morgenstern.
\newblock {\em Physica}, {\bf C153--155}, 59 (1988).

\bibitem{PUE88}
B.~P{\"u}mpin, H.~Keller, W.~K{\"u}ndig, W.~Odermatt, B.D. Patterson, J.W.
  Schneider, H.~Simmler, S.~Connell, K.A. M{\"u}ller, J.G. Bednorz, K.W.
  Blazey, I.~Morgenstern, C.~Rossel, and I.M. Savic.
\newblock {\em Z. Phys.}, {\bf B72}, 175 (1988).

\bibitem{MOR88/5}
I.~Morgenstern, K.A. M{\"u}ller, and J.G. Bednorz.
\newblock {\em Physica}, {\bf B152}, 85 (1988).

\bibitem{KEL88}
H.~Keller, P.~P{\"u}mpin, W.~K{\"u}ndig, W.~Obermatt, B.D. Patterson, J.W.
  Schneider, K.A. M{\"u}ller, J.G. Bednorz, K.W. Blazey, I.~Morgenstern,
  C.~Rossel, and I.M. Savic.
\newblock {\em Physica}, {\bf C153--155}, 71 (1988).

\bibitem{MOR89/2}
I.~Morgenstern.
\newblock {\em IBM Journal of Research and Development}, {\bf 33}, 307 (1989).

\bibitem{ROS89}
C.~Rossel, Y.~Maeno, and I.~Morgenstern.
\newblock {\em Phys.\ Rev.\ Letters}, {\bf 62}, 681 (1989).

\bibitem{EBN85}
C.~Ebner and D.~Stroud.
\newblock {\em Phys.\ Rev.}, {\bf B31}, 165 (1985).

\bibitem{TSU98}
C.C. Tsuei and T.~Doderer.
\newblock `Charge confinement effect in cuprate superconductors -- an explanation
for the normal-state resistivity and pseudogap',
\newblock {\em preprint}, 1998.

\bibitem{Tranq}
J.M. Tranquada, J.D. Axe, N~Ichikawa, A.R. Moodenbaugh, Y.~Nakamura,
and S.~Uchida.
\newblock {\em Phys.\ Rev.\ Letters}, {\bf 78}, 338 (1997).

\bibitem{Bianc}
N.L. Saini, A.~Lanzara, H.~Oyanagi, H.~Yamaguchi, K.~Oka, T.~Ito, and
A.~Bianconi.
\newblock {\em Phys.\ Rev.}, {\bf B55}, 12759 (1997).

\bibitem{CHE89}
C.J. Chen and C.C. Tsuei.
\newblock {\em Solid State Communication}, {\bf 71}, 33 (1989).

\bibitem{MUE87}
K.A. M{\"u}ller, M.~Takashige, and J.G. Bednorz.
\newblock {\em Phys.\ Rev.\ Letter}, {\bf 58}, 1143 (1987).

\bibitem{SIG92}
M.~Sigrist and T.M. Rice.
\newblock {\em J. Phys.\ Soc.\ Jpn.}, {\bf 61}, 4283 (1992).

\bibitem{MOR94}
I.~Morgenstern, W.~Fettes, T.~Husslein, C.~Baur, H.-G. Matuttis, and J.M.
  Singer.
\newblock {\em Proc.\ PC94 Conference, Lugano}, 1994.

\bibitem{HUS94}
T.~Husslein, I.~Morgenstern, D.M. Newns, P.C. Pattnaik, and J.M. Singer.
\newblock {\em Phys. Rev}, {\bf B54}, 16179 (1996), and references therein.

\bibitem{AND87}
P.W. Anderson.
\newblock {\em Science}, {\bf 235}, 1196 (1987).

\bibitem{TSU94}
C.C. Tsuei, J.K. Kirley, C.C. Chi, L.S. Yu-Jahnes, A.~Gupta, T.~Shaw, J.Z. Sun,
  and M.B. Ketchen.
\newblock {\em Phys.\ Rev.\ Lett.}, {\bf 73}, 593 (1994).

\bibitem{SCA89}
R.T. Scalettar, E.Y. Loh, J.E. Gubernatis, A.~Moreo, S.R. White, D.J.
  Scalapino, R.L. Sugar, and E.~Dagotto.
\newblock {\em Phys.\ Rev.\ Lett.}, {\bf 62}, 1407 (1989).


\bibitem{FRI91}
M.~Frick, I.~Morgenstern, and W.~{von der Linden}.
\newblock {\em Z. Phys.}, {\bf B82}, 339 (1991).

\bibitem{YAN62}
C.N. Yang.
\newblock {\em Rev.\ Mod.\ Phys.}, {\bf 34}, 694 (1962).

\bibitem{SUG86}
G.~Sugiyama and S.E. Koonin.
\newblock {\em Ann.\ Phys.}, {\bf 168}, 1 (1986).

\bibitem{SOR89}
S.~Sorella, A.~Parola, M.~Parrinello, and E.~Tosatti.
\newblock {\em Int.\ J. Mod.\ Phys.}, {\bf B3}, 1875 (1989).

\bibitem{HUS97}
T.~Husslein, W.~Fettes, and I.~Morgenstern.
\newblock {\em Int.\ J. Mod.\ Phys.}, {\bf C8}, 397 (1997).

\bibitem{FET98/3}
W.~Fettes.
\newblock Ph.D. Thesis, Universit{\"a}t Regensburg, 1998.

\bibitem{FET97/3}
W.~Fettes and I.~Morgenstern.
\newblock {\em accepted by European Physical Journal B}, 1998.

\bibitem{HIR88}
J.E. Hirsch.
\newblock {\em Phys.\ Rev.}, {\bf B38}, 12023 (1988).

\bibitem{WHI89}
S.R. White, D.J. Scalapino, R.L. Sugar, and N.E. Bickers.
\newblock {\em Phys.\ Rev.\ Lett.}, {\bf 63}, 1523 (1989).

\bibitem{IMA91}
M.~Imada.
\newblock {\em J.\ Phys.\ Soc.\ Jpn.}, {\bf 60}, 2740 (1991).

\bibitem{ZHA97}
S.~Zhang, J.~Carlson, and J.E. Gubernatis.
\newblock {\em Phys.\ Rev.\ Lett.}, {\bf 78}, 4486 (1997).

\bibitem{YAM98}
K. Yamaji, T. Yanagisawa, T. Nakanishi, and S. Koike,
\newblock preprint, 1998.

\bibitem{HUS96}
T.~Husslein.
\newblock Ph.D. Thesis, Universit{\"a}t Regensburg, 1996.

\bibitem{FET98}
W.~Fettes and I.~Morgenstern.
\newblock {\em Int. J. Mod. Phys.}, {\bf C9}, 943 (1998).

\bibitem{FET97/2}
W.~Fettes, I.~Morgenstern, and T.~Husslein.
\newblock {\em Int. J. Mod. Phys.}, {\bf C8}, 1037 (1997).

\bibitem{BLA81}
R.~Blankenbecler, D.J. Scalapino, and R.L. Sugar.
\newblock {\em Phys.\ Rev.}, {\bf D24}, 2278 (1981).

\bibitem{DIN96}
H.~Ding, T.~Yokoya, J.C. Campuzano, T.~Takahashi, M.~Randeria, M.R. Norman,
T.~Mochiku, K.~Kadowaki, and J.~Giapintzakis.
\newblock {\em Nature}, {\bf 382}, 51 (1996).

\bibitem{LOE96}
A.G. Loeser, Z.-X. Shen, D.S. Dessau, D.S. Marshall, C.H. Park, P.~Fournier,
and A.~Kapitulnik.
\newblock {\em Science}, {\bf 273}, 325 (1996).

\bibitem{LOE96/2}
A.G. Loeser, D.S. Dessau, and Z.-X. Shen.
\newblock {\em Physica}, {\bf C263}, 208 (1996).

\bibitem{HAR96}
J.M. Harris, Z.-X. Shen, P.J. White, D.S. Marshall and M.C. Schabel,
J.N. Eckstein, and I. Bozovic.
\newblock {\em Phys.\ Rev.}, {\bf B54}, R15665 (1996).

\bibitem{KELpriv97}
H.~Keller, D.~Brinkman, and T.~Schneider.
\newblock private communication, 1997.

\bibitem{HYB89}
M.S. Hybertsen, M. Schl{\"u}ter, and N.E. Christensen.
\newblock {\em Phys.\ Rev.}, {\bf B39}, 9028 (1989).

\bibitem{BAC91}
S.B. Bacci, E.R. Gagliano, R.M. Martin, and J.F. Annett.
\newblock {\em Phys.\ Rev.}, {\bf B44}, 7504 (1991).

\bibitem{SON95}
J. Song and J.F. Annett.
\newblock {\em Phys.\ Rev.}, {\bf B51}, 3840 (1995).

\bibitem{DENpriv}
D.M. Newns.
\newblock private communication, 1998.

\bibitem{FET95}
W.~Fettes, I.~Morgenstern, T.~Husslein, H.-G. Matuttis, J.M. Singer, and
  C.~Baur.
\newblock {\em J. Phys.\ I France}, {\bf 5}, 455 (1995).

\bibitem{KELpriv}
H.~Keller.
\newblock private communication, 1997.

\bibitem{KIV98}
O.~Zachar, S.A. Kivelson, and V.J. Emery.
\newblock {\em Phys.\ Rev.}, {\bf B57}, 1422 (1998).

\end{thebibliography}


\end{document}